\documentclass[pre,preprint,showpacs,preprintnumbers,amsmath,amssymb]{revtex4}
\usepackage{graphics,epstopdf,epsfig,stmaryrd}

\newcommand{\mI}{\mathcal{I}}
\newcommand{\mE}{\mathcal{E}}
\newcommand{\mJ}{\mathcal{J}}
\newcommand{\mA}{v_{tr}}

\newcommand{\dvdv}{dv_{\phi}/dv_{tr}}
\begin {document}
\title{Nonlocal adiabatic theory. I. The action distribution function}
\author{Didier B\'enisti}
\email{didier.benisti@cea.fr} 
\affiliation{ CEA, DAM, DIF F-91297 Arpajon, France.}
\date{\today}
\begin{abstract}
In this paper, we address the motion of charged particles acted upon by a sinusoidal electrostatic wave, whose amplitude and phase velocity vary slowly enough in time for neo-adiabatic theory to apply. Moreover, we restrict to the situation when only few separatrix crossings have occurred, so that the adiabatic invariant, $\mI$, remains nearly constant.  We insist here on the fact that $\mI$ is different from the dynamical action, $I$. In particular, we show that $\mI$ depends on the whole time history of the  wave variations, while the action is usually defined as a local function of the wave amplitude and phase velocity. Moreover, we provide several numerical results showing how the action distribution function, $f(I)$, varies with time, and we explain how to derive it analytically. The derivation is then generalized to the situation when the wave is weakly inhomogeneous. 
\end{abstract}
\maketitle
\section{Introduction}
Nonlinear interactions of charged particles with electrostatic waves are ubiquitous in plasma physics and have many important implications. To cite a few examples that motivated the present work, nonlinear wave-particle interactions may completely change the growth and saturation levels of stimulated Raman scattering (SRS), compared to linear predictions. This has been shown experimentally in Ref.~\cite{mont}, and reproduced numerically in Refs.~\cite{yin,SRS3D}. Moreover, unexpected high levels of Raman reflectivity, measured during the National Ignition Campaign at Livermore~\cite{NIC}, might also be due to the nonlinear electron response to the SRS-driven plasma wave. For the parameters relevant to laser-plasma interaction in a fusion device (i.e., laser intensity, plasma density, electron temperature, \dots) the SRS growth rate is small enough for the electron motion to be considered as adiabatic. In particular, it has been shown in Refs.~\cite{benisti07,benisti08,lindberg}~that an adiabatic estimate of the nonlinear frequency shift of the  SRS-driven plasma wave was in excellent agreement with numerical results (as those given by direct Vlasov simulations of SRS in Ref.~\cite{benisti08}). Moreover, in the companion paper~\cite{companion}, we explicitly make use of the distribution function obtained in this article in order to generalize the derivation of the frequency shift by accounting for plasma inhomogeneity and multidimensional effects. 

Although we assume that the particles are only acted upon by an electrostatic wave, the results derived in this paper are relevant to magnetized plasmas. Indeed, in Ref.~\cite{breizman}, the motion of adiabatically trapped particles in an electrostatic wave with sweeping frequency is analyzed in order to provide a nonlinear description of tokamak instabilities, such as nonlinear Alfven waves excited by fast ions. Moreover, in space plasmas, all kinds of energetic particles result from their interactions with electrostatic waves. In some instances, the energization mechanism strongly relies on the presence of a magnetic field (see for example Refs.~\cite{MIT1,MIT2,MIT3}). However, as argued recently in Ref.~\cite{mourenas}, energetic electrons in the inner magnetosphere may result from adiabatic trapping in electrostatic whistler modes. Then, although the electron motion may be a bit more complicated than that investigated in this paper due to the presence of a magnetic field, the time evolution of the electron distribution function seems to be essentially due to the very same adiabatic trapping and detrapping mechanisms as those discussed in the present article~\cite{mourenas2}.

\begin{figure}[!h]
\centerline{\includegraphics[width=10cm]{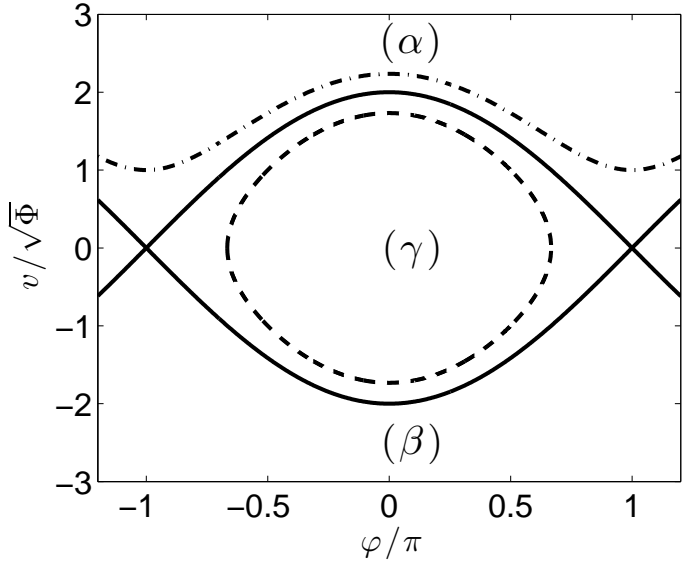}}
\caption{\label{f1} The separatrix (solid line), a frozen trapped orbit (dashed line) and a  frozen untrapped orbit (dashed-dotted line) for the dynamics of Hamiltonian $H$, defined by Eq.~(\ref{H}). The regions $(\alpha)$, $(\beta)$ and $(\gamma)$ are labelled.}
\end{figure}

As is well known, adiabatic theory applies to the nearly periodic and slowly varying dynamics of a Hamiltonian, $H(\varphi,v,\varepsilon t)$, with $\varepsilon \ll 1$. Here,  $\varphi$ and $v$ are canonically conjugated variables, and $H$ is periodic in $\varphi$, so that the frozen orbits  [the orbits of $H(\varphi,v,\varepsilon t_0)$ where $t_0$ is a constant] are periodic. Then, if the largest time period of a frozen orbit is of the order of unity, the action,
\begin{equation}
\label{1}
I \equiv \frac{1}{2\pi}Ê\oint v d\varphi,
\end{equation}
where the integral is calculated along a frozen orbit, is an adiabatic invariant, its variations are of the order of $\varepsilon$ within a time interval of the order of $1/\varepsilon$~\cite{lenard,arnold} (and one might even introduce better conserved invariants, but they would not be useful within the context of this paper~\cite{note}). Now, it is also well-known that, if the phase portrait contains a separatrix (i.e., a frozen orbit with an unstable fixed point), the aforementioned theorem does not apply because the time period is infinite on the separatrix. Nevertheless, following the so-called neo-adiabatic theory, several authors~\cite{tim,cary,hanna,nei,ten,vas} proved that the adiabatic invariant remained nearly conserved, and only changed by an amount of the order of $\varepsilon \ln(\varepsilon)$ due to separatrix crossing. However, the adiabatic invariant no longer is the action, $I$, as defined by Eq.~(\ref{1}). Indeed, for the dynamics considered in this paper, the phase space may be divided into three regions, region $(\alpha)$ above the upper branch of the separatrix, region $(\beta)$ below the lower branch of the separatrix, and region $(\gamma)$ inside the separatrix (see Fig.~\ref{f1}). Then,  $I$ is only continuous within each region, and each time an orbit moves from one region to another, it experiences a jump, that is purely geometric. 
Such a geometrical change in the action has often been considered as inessential and overlooked, while it actually has important physical implications. Indeed, the action $I$, as defined by Eq.~(\ref{1}), is purely local in the wave amplitude and phase velocity. Consequently, if it could be considered as a true invariant, the corresponding adiabatic distribution function would also be local (see for example Ref.~\cite{krapchev}, and the later discussion in Ref.~\cite{DNC}), which would imply that adiabatic dynamics should be reversible. However, as pointed out by several authors, these dynamics are usually irreversible, precisely because of separatrix crossings. For example, it has been shown in Refs.~\cite{vgroup,dissipation} that adiabatic trapping by an electrostatic wave entailed an irreversible increase in kinetic energy and, therefore, wave dissipation. Such a dissipation manifests itself by the shrinking of the wave packet, both along and across its direction of propagation. Moreover, as discussed in Ref.~\cite{ratchet}, an electrostatic wave with constant amplitude and slowly oscillating phase velocity would entail a directed flow, thus leading to a so-called ratchet effect. 

Now, from the action $I$, one may actually derive a true adiabatic invariant, $\mI$, precisely defined by Eq.~(\ref{35}) of Paragraph~\ref{II.2}. In this definition for $\mI$ enter explicitly  the values assumed by the wave phase velocity each time the considered orbit has crossed the separatrix. Hence, $\mI$ depends on the whole history of the wave evolution.~Consequently, when the dynamics is assumed to be adiabatic (i.e., when $\mI$ is assumed to be a constant of motion), the particles distribution function is usually nonlocal, and its time variation is usually irreversible. Although such a nonlocality has already been discussed in the past, we could not find any reference that explicitly derived the adiabatic distribution function in a simple way. The present article aims at filling this gap. Instead of working with the true adiabatic invariant, $\mI$, we prefer deriving the action distribution function, $f(I)$, to show how it evolves in time in the adiabatic regime. 

A crucial step in the derivation of $f(I)$ is the use of previously published results~\cite{cary,neishtadt} on  transition probabilities from one region of phase space to another one. In this article, we summarize the main steps of the derivation within a few lines, where we point out that the corresponding results only apply to distributions in action that are smooth enough. This is illustrated by several numerical simulations, which are also used to test the accuracy of adiabatic predictions as regards separatrix crossing. Moreover, starting from an initial Dirac distribution in action, we numerically show that $f(I)$ evolves towards of sum of Dirac distributions, as is clear from Figs.~\ref{f6}(d) and~\ref{f8}(d) of Paragraph~\ref{III.3}. Hence, we provide unambiguous numerical evidences of the non-conservation of the action distribution function. 

Actually, the considered dynamics may be represented by two different Hamiltonians, leading to two different definitions for the action. In the adiabatic regime, we show that these two different actions are very close to each other, leading to similar predictions as regards the transitions from one region of phase space to another. Nevertheless, as  may be inferred theoretically, and confirmed numerically, one Hamiltonian formulation systematically leads to more accurate predictions, in the adiabatic regime. However, the other formulation is useful in a regime that is slightly beyond the adiabatic one, to predict when trapping is purely impossible, even when the wave grows.

Finally, based on adiabatic results, we explicitly derive how the action distribution function changes due to separatrix crossings, which is our main result. This result is then generalized to the situation when the wave amplitude and phase velocity vary in space and time. In this situation, there is no adiabatic invariant for the untrapped particles. Consequently, in addition to the jumps in $f(I)$ due to separatrix crossings, one must also account for the continuous change in $f(I)$ for the passing particles, which we also derive. 

The article is organized as follows. In Section~\ref{II}, we introduce the dynamical system we focus on, and we show that it may be correctly represented by at least two Hamiltonians. Then, we show how to construct an adiabatic invariant for each of these Hamiltonians, and discuss their relative merits. In Section~\ref{III}, we first quickly remind the main steps that lead to the derivation of the probability transitions from one region of phase space to another one. Then we present several  simulation results that test the accuracy of adiabatic theory. These simulations unambiguously show that the action distribution function is not conserved. Using the results of Sections~\ref{II}~and~\ref{III}, we provide in Section~\ref{IV}~an explicit way to derive the action distribution function in the adiabatic regime, i.e., when the Hamiltonian is nearly periodic, slowly varying, and when only few separatrix crossings occurred. These results are then briefly generalized in Section~\ref{V} to the situation when the wave is inhomogeneous, so that there is no adiabatic invariant for the passing particles. Section~\ref{VI}~summarizes and concludes our work. 

\section{Hamiltonian formulations for the dynamics, and adiabatic invariants}
\label{II}
\subsection{Hamiltonian formulations}
\label{II.1}
In this paper, we focus on the one-dimensional dynamics of charged particles, acted upon by an electrostatic wave whose electric field reads,
\begin{equation}
\label{2}
\mE(x,t)=E_0 \sin[\varphi(x,t)],
\end{equation}
and except in Section~\ref{V}, the wave amplitude, $E_0$, only depends on time. The wave number, $k$, and frequency, $\omega$, are defined by $k\equiv \partial_x \varphi$, $\omega=-\partial_t \varphi$ and, except in Section~\ref{V}, they are space-independent. Moreover, we henceforth assume that $E_0$, $k$ and $\omega$ vary slowly with time. From $k$ and $\omega$ we introduce the wave phase velocity, $V_\phi \equiv \omega/k$. Then, for particles with charge $Q$ and mass $M$, the equations of motion are,
\begin{eqnarray}
\label{3}
dx/dt & =& p/M, \\
\label{4}
dp/dt&=&QE_0 \sin[\varphi(x,t)].
\end{eqnarray}
They may be derived from the following Hamiltonian,
\begin{equation}
\label{5}
H_1=p^2/2M+(QE_0/k) \cos[\varphi(x,t)],
\end{equation}
for the conjugated variable $x$ and $p$. Now, even though $E_0$, $k$ and $\omega$ vary slowly in time, $\omega$ is not necessarily small, so that $\varphi$ may rapidly vary with time. Then, $H_1$ is not necessarily slowly varying, and the adiabatic theory does not apply to $H_1$. 

Actually, for adiabatic results to apply, one must use $\varphi$ as one of the canonically conjugated variables. Then, let us introduce some velocity scale $v_{th}$ (whose value will be discussed in a few lines), and the dimensionless time, $\tau$, so that $d\tau=k v_{th}dt$. The equations of motion now read,
\begin{eqnarray}
\label{6}
d\varphi/d\tau & =& v-v_\phi, \\
\label{7}
dv/d\tau&=&-\Phi \sin(\varphi),
\end{eqnarray}
where $v \equiv v_{th}^{-1} (dx/dt)$, $v_\phi \equiv  V_{\phi}/v_{th}$, and where we have denoted, 
\begin{equation}
\label{n1}
\Phi \equiv -QE_0/kT_e,
\end{equation}
with $T_e\equiv Mv_{th}^2$.   Eqs.~(\ref{6}) ~and~(\ref{7}) derive from the following Hamiltonian,
\begin{equation}
\label{H}
H=Ê(v-v_\phi)^2/2-\Phi \cos(\varphi),
\end{equation}
 where, now, $v$ and $\varphi$ are the canonically conjugated variables. 
 
 For Hamiltonian $H$, the slow time scale, which we denote by $T_w$, is related to the time evolution of the wave properties, namely the time evolution of $\Phi$ and of $v_\phi$. The fast time scale is  is related to the evolution of $\varphi$ and, as usual, it is the typical period, $T$, of a frozen orbit (away from the separatrix). 
 
The conditions for $T_w$ and  $T$ to be indeed slow and fast are discussed in Appendix~\ref{A0}. When the initial distribution in velocity is a Dirac distribution, so that all particles have the same velocity $p_0/M$ when the wave amplitude is vanishingly small, the condition reads,
\begin{equation}
\label{n2}
\varepsilon \equiv 1/(v_0-v_\phi)T_w\ll 1,
\end{equation}
where $v_0 \equiv p_0/Mv_{th}$, and where $v_\phi$ is the normalized wave phase velocity when the particles are just about to be trapped. Then, for an initial Dirac distribution in velocity, a good choice for $v_{th}$ is $v_{th} = p_0/M-V_\phi$, where  $V_\phi$ is one value of the wave phase velocity,  of the order of that corresponding to separatrix crossing. Such a choice for $v_{th}$ has been explicitly made in Ref.~\cite{action}. It is not systematically imposed in this paper, since we rather let $v_0$ be a free parameter. However, $v_0-v_\phi$ is always chosen to be of the order of unity. 

For a smooth initial velocity distribution function, the condition for $T_w$ and $T$ to be indeed slow and fast reads,
\begin{equation}
\label{n3}
\varepsilon_S \equiv 1/v_TT_w\ll 1,
\end{equation}
where $v_T$ is the typical range of variation of the initial distribution function, about the wave phase velocity. However, there are restrictions regarding the applicability of Eq.~(\ref{n3}), which are discussed in Appendix~\ref{A0}. For an initial Maxwellian distribution function, $v_T$ is just the thermal velocity, which appears as the natural choice for $v_{th}$. Then, in Eq.~(\ref{n1}), $T_e$ is the particles' temperature. 

In most of the paper, we focus on initial Dirac distributions in velocity. Then, since $H$ is periodic in $\varphi$ and varies slowly in time when Eq.~(\ref{n2}) is fulfilled, the neo-adiabatic results of Refs.~\cite{tim,cary,hanna,nei,ten,vas} directly apply to $H$. Namely, the adiabatic invariant changes by a term of the order $\varepsilon \ln(\varepsilon)$ due to separatrix crossing. This has been checked numerically in Ref.~\cite{action}, together with the accuracy of neo-adiabatic predictions depending on how small $\varepsilon$ is (and the change in the adiabatic invariant when $\varepsilon$ is not small has also been derived analytically in Ref.~\cite{action}).\\
 
 Alternatively, one may want to use $v_2 \equiv v-v_\phi$ and $\varphi$ as conjugated variables. Then
 \begin{eqnarray}
\label{8}
d\varphi/d\tau & =& v_2, \\
\label{9}
dv_2/d\tau&=&-\Phi \sin(\varphi)-\dot{v}_\phi,
\end{eqnarray}
where $\dot{v}_\phi \equiv dv_\phi/d\tau$. Clearly, Eqs.~(\ref{8})~and~(\ref{9})~derive from the Hamiltonian,
\begin{equation}
\label{H2}
H_2=v_2^2/2-\Phi \cos(\varphi)+\varphi \dot{v}_\phi.
\end{equation}
For $H_2$, the fast time scale is still associated with the typical period, $T$, of a frozen orbit (away from the separatrix) and the slow time scale, $T_w$, is associated with the evolution of the wave properties, i.e., the time evolution of $\Phi$ and $\dot{v}_\phi$ (and not $v_\phi$ as for $H$). When Eq.~(\ref{n2}) or Eq.~(\ref{n3}) is fulfilled, $H_2$ may be considered as slowly varying.  However, it is not periodic in $\varphi$, so that neo-adiabatic results do not directly apply to $H_2$. Nevertheless, as discussed in Paragraph~\ref{II.2.2}, one may define an adiabatic invariant from $H_2$.

\subsection{Adiabatic invariant for H}
\label{II.2}

For the dynamics defined by $H$, we actually use a slightly different definition for the action, whether the orbit lies in region $(\gamma)$, or in region $(\alpha)$ or $(\beta)$. For the so-called untrapped particles [those whose orbit lies either in region $(\alpha)$ or in region $(\beta)$], we keep the definition Eq.~(\ref{1}) for $I$, namely,
\begin{equation}
\label{10}
I \equiv \frac{1}{2\pi} \oint v d\varphi,
\end{equation}
where the integral is calculated along a frozen untrapped orbit (as that plotted in Fig.~\ref{1}) i.e., along an orbit corresponding to a constant value of $H$. Note that the frozen dynamics of $H$ is 2$\pi$-periodic, so that the phase space is topologically equivalent to a cylinder and, on the cylinder, the orbits in region $(\alpha)$ and $(\beta)$ are closed. This property is explicitly used in the definition Eq.~(\ref{10}) of the action. The integral in Eq.~(\ref{10}) is to be understood as an integral over a closed orbit on the cylinder. 

From Eq.~(\ref{7}), $d\varphi$ has the same sign as $v-v_\phi$. Therefore, it is positive in region $(\alpha)$ and negative in region $(\beta)$. Hence, the action in region $(\alpha)$ which we denote by $I_\alpha$ reads,
\begin{equation}
\label{11}
I_\alpha= \frac{1}{2\pi} \oint (v-v_\phi) d\varphi +v_\phi,
\end{equation}
while in region $(\beta)$, the action which we denote by $I_\beta$ reads,
\begin{equation}
\label{12}
I_\beta =  \frac{1}{2\pi} \oint (v-v_\phi) d\varphi -v_\phi.
\end{equation}
From Eqs.~(\ref{11}) and (\ref{12}) it is clear that one cannot use a uniform definition for the action valid in both regions $(\alpha)$ and $(\beta)$. For the same value of $\oint (v-v_\phi)d\varphi$, $I_\alpha$ and $I_\beta$ differ by $\pm 2v_\phi$. 

For a trapped particle, whose orbit lies in region $(\gamma)$, we define its action by
\begin{eqnarray}
\nonumber
I_\gamma &=& \frac{1}{4\pi} \oint v d\varphi \\
\label{13}
&=&\frac{1}{4\pi} \oint (v-v_\phi) d\varphi,
\end{eqnarray}
where the integral is calculated along a frozen trapped orbit, as that plotted in Fig.\ref{1}, so that along such an orbit it is clear that $\oint d\varphi=0$. The choice to divide the integral by $4\pi$ instead of $2\pi$ is vindicated by the fact that, close to the separatrix, $\oint (v-v_\phi) d\varphi$ is twice as large along a trapped orbit than along an untrapped one. Then, adopting the definition Eq.~(\ref{13}) for $I_\gamma$ actually simplifies analytical calculations, as those performed in the companion paper~\cite{companion} for the derivation of the nonlinear frequency shift. 

In order to clarify the geometrical changes in the action, let us introduce, 
\begin{eqnarray}
\label{24}
m&=& 2\Phi/(H+\Phi) \\
\label{25}
v_{tr} &=& 4\sqrt{\Phi}/\pi,
\end{eqnarray}
then, it is well-known (see for example Ref.~\cite{benisti07}) that, 
\begin{eqnarray}
\label{26}
I_\alpha&=& v_{tr} E(m)/\sqrt{m}+v_\phi \\
\label{27}
I_\beta&=& v_{tr} E(m)/\sqrt{m}-v_\phi \\
\label{28}
I_\gamma&=& v_{tr} [E(m^{-1})+(m^{-1}-1)K(m^{-1})],
\end{eqnarray}
where $K(m)$ and $E(m)$ are, respectively, the Jacobian elliptic integrals of first and second kind~\cite{abramowitz}. Now, from adiabatic theory we know that, as long as a particle orbit remains in region $(\alpha)$, its action is nearly conserved, $I_\alpha \approx I_\alpha(0)$ [moreover, if the initial wave amplitude is vanishingly small, it is clear from Eq.~(\ref{10}) that $I_\alpha(0)$ is nothing but the initial particle velocity, $v_0$]. Moreover, since $E(m)/\sqrt{m}Ê\geq 1$, is clear that Eq.~(\ref{26}) may only be fulfilled when $v_{tr}+v_\phi \leq I_\alpha(0)$. Therefore, the orbit has to leave region $(\alpha)$ when
\begin{equation}
\label{49}
 I_\alpha=v_{tr}+v_\phi\text{ and }\dot{v}_{tr}+\dot{v}_\phi>0.
\end{equation}
Geometrically speaking, this happens when the orbit is very close to the upper branch of the separatrix, and when this upper branch keeps moving upward in phase space. Now, the conditions $I_\alpha=v_{tr}+v_\phi$ is only fulfilled when $m=1$. Moreover, from neo-adiabatic theory we know that, after separatrix crossing, the value of $m$ is nearly conserved. Hence, if the orbit moves to region ($\beta$) then, from Eq.~(\ref{27}) with $m=1$, one finds that its action becomes $I_\beta=v_{tr}-v_\phi$, so that there is a jump in the particle's action by,
\begin{equation}
\label{29}
\Delta I_{\alpha \rightarrow \beta}=-2v_{\phi_{\alpha \rightarrow \beta}},
\end{equation}
where $v_{\phi_{\alpha \rightarrow \beta}}$ is a constant, it is the value assumed by the wave phase velocity when the transition from region $(\alpha)$ to region $(\beta)$ occurred. From Eq.~(\ref{29}), it is clear that the action is not conserved. The jump in action calculated in Eq.~(\ref{29}) [and in Eqs.~(\ref{30})-({32})] is purely geometric, i.e., it corresponds to the change in action in the limit $\varepsilon \rightarrow 0$.  

Similarly, if the orbit moves from region $(\alpha)$ to region $(\gamma)$, from Eqs.~(\ref{26}) and (\ref{28}) with $m=1$ one easily finds that the action changes by,
\begin{equation}
\label{30}
\Delta I_{\alpha \rightarrow \gamma}=-v_{\phi_{\alpha \rightarrow \gamma}},
\end{equation}
where $v_{\phi_{\alpha \rightarrow \gamma}}$ is the value assumed by the wave phase velocity when the transition from region $(\alpha)$ to region $(\gamma)$ occurred.

Using the same kind of reasoning and notations, one easily finds that a particle orbit necessarily leaves region $(\beta)$ if,
\begin{equation}
\label{50}
I_\beta=v_{tr}-v_\phi\text{ and }\dot{v}_{tr}-\dot{v}_\phi>0,
\end{equation}
which happens when the orbit is very close to the lower branch of the separatrix and when this lower branch keeps moving downward in phase space. Then, the orbit may end up either in region $(\alpha)$ or in region $(\gamma)$, leading to the respective jumps in action,
\begin{eqnarray}
\label{31}
\Delta I_{\beta \rightarrow \alpha}&=&+2v_{\phi_{\beta \rightarrow \alpha}}, \\
\label{32}
\Delta I_{\beta \rightarrow \gamma}&=&+v_{\phi_{\beta \rightarrow \gamma}}.
\end{eqnarray}

Similarly, the orbit necessarily leaves region $(\gamma)$ when,
\begin{equation}
\label{51b}
I_\gamma= v_{tr}\text{ and }\dot{v}_{tr}<0.
\end{equation}
It may either move to region $(\alpha)$ or to region $(\beta)$, leading to the respective jumps in action,
\begin{eqnarray}
\label{33}
\Delta I_{\gamma \rightarrow \alpha}&=&+v_{\phi_{\gamma \rightarrow \alpha}}, \\
\label{34}
\Delta I_{\gamma \rightarrow \beta}&=&-v_{\phi_{\gamma \rightarrow \beta}}.
\end{eqnarray}

Let us now introduce the dynamical variable $\mI$ defined by, 
\begin{equation}
\label{35}
\mI = I(\tau)-2\sum_{n_1} v_{\phi_{\beta \rightarrow \alpha}}^{(n_1)}+2\sum_{n_2} v_{\phi_{\alpha \rightarrow \beta}}^{(n_2)}-\sum_{n_3} v_{\phi_{\gamma \rightarrow \alpha}}^{(n_3)}+\sum_{n_4} v_{\phi_{\gamma \rightarrow \beta}}^{(n_4)}-\sum_{n_5} v_{\phi_{\beta \rightarrow \gamma}}^{(n_5)}+\sum_{n_6} v_{\phi_{\alpha \rightarrow \gamma}}^{(n_6)},
\end{equation}
where the sum is over all the transitions from one region of phase space to another one experienced by the considered particle's orbit before time $\tau$, and where $v_{\phi_{\xi \rightarrow \zeta}}^{(n)}$ is the value assumed by the wave phase velocity when the transition from region $(\xi)$ to region $(\zeta)$ occurred for the $n^{th}$ time. Then from Eqs.~(\ref{26})-(\ref{34}), it is clear that the geometrical changes in the action have been absorbed in the definition for $\mI$, so that $\mI$ is a genuine adiabatic invariant. Hence, if we denote by $\mathcal{F}$ the distribution in $\mI$, $\mathcal{F}(\mI,t) \approx \mathcal{F}(\mI,0)$. Now, $\mI$ depends on the whole time history of the wave amplitude and  phase velocity, so that it is nonlocal in the wave parameters. Consequently, the particle's adiabatic distribution function is also nonlocal. 

Now, using $\mI$ as a dynamical variable is not convenient. Indeed, it requires the knowledge of all the $v_{\phi_{x\rightarrow y}}^{(n)}$'s. Therefore, it may only be defined up to a given time, and may only be defined in advance if the time evolution of the wave amplitude and phase velocity are also known in advance, which is not the case if one solves self-consistently for the evolution of the wave and of the particles. Moreover, the definition for $\mI$ is rather involved and not easy to manipulate. Consequently, we will henceforth only consider the action $I$, and derive its distribution function, $f(I)$. \\

\subsection{Adiabatic invariant for Hamiltonian $H_2$}
\label{II.2.2}
As discussed in Paragraph~\ref{II.1}, $H_2$ seems less suited than $H$ to apply adiabatic results. Nevertheless, several authors (see for example Refs.~\cite{mourenas2,armon}) recently used Hamiltonian $H_2$ to investigate trapping probabilities in the adiabatic limit. Therefore, we found it useful to compare the predictions made using $H_2$ with those derived from $H$. 

Again, we divide phase space into three distinct regions. Region $(\alpha)$ is associated with passing particles such that $v_2>0$, region $(\beta)$ is associated with passing particles such that $v_2<0$, and region $(\gamma)$ is associated with trapped particles. For the latter class of particles, we use a definition for the action which is very similar to that used for $H$, namely we define,
\begin{equation}
\label{44}
I_{2} \equiv \frac{1}{4\pi} \oint v_2 d\varphi,
\end{equation}
where the integral is over a frozen trapped orbit, so that $I_{2}$ is only a function of $H_2$ and, reciprocally, $H_2 \equiv H_2(I_{2})$. Then, $I_{2} \approx Const.$ for trapped particles.

As regards the untrapped particles, the situation is more complicated because their frozen orbits are not closed, since the $H_2$ is not periodic. Then, for these orbits, we define,
\begin{equation}
\label{45}
I_2(H_2,\varphi) \equiv  \frac{1}{2\pi} \int_{\varphi-\eta \pi}^{\varphi+\eta \pi} v_2 d\varphi',
\end{equation}
where $\eta$ is the sign of $v_2$. The latter value of $I_2$ is not constant, because it explicitly depends of $\varphi$. In order to derive its time evolution we recall that the adiabatic result for $I$, which guarantees its near conservation, just amounts to proving that $dI/d\tau \approx \langle dI/d\tau\rangle$, where $\langle . \rangle$ stands for the space averaging over a frozen orbit~\cite{arnold,benisti16}. Then we estimate,
\begin{eqnarray}
\label{43b}
\frac{dI_2}{d\tau} & \approx & \left \langle \frac{dI_2}{d\tau}Ê\right \rangle \\
\label{48}
&\approx&-\eta \dot{v}_\phi.
\end{eqnarray}
Therefore, for passing particles, the action should be defined as $I_2+\eta v_\phi$, and not as $I_2$, since within region~$(\alpha)$ or $(\beta)$, $I_2+\eta v_\phi$ remains nearly constant. 

In conclusion, let us introduce, for a passing orbit,
\begin{equation}
\label{14}
J(\varphi,H_2) \equiv \frac{1}{2\pi}\int_{\varphi - \pi}^{\varphi + \pi}Ê\sqrt{2[H_2+\Phi \cos(\varphi')-\varphi' \dot{v}_\phi]} d\varphi',
\end{equation}
and for a trapped orbit,
\begin{equation}
\label{50b}
J_t(H_2) \equiv \frac{1}{4\pi} \intÊ\sqrt{2[H_2+\Phi \cos(\varphi)-\varphi \dot{v}_\phi]} d\varphi,
\end{equation}
where the integral is over the values of $\varphi$ (spanning an interval whose amplitude is less than $2\pi$) leaving $[H_2+\Phi \cos(\varphi)-\varphi \dot{v}_\phi]$ positive. Let us moreover define the dynamical variable $J_2$, respectively in region $(\alpha)$, $(\beta)$ and $(\gamma)$, by
\begin{eqnarray}
\label{51}
J_{2\alpha} &\equiv &J +v_\phi, \\
\label{52}
J_{2\beta} & \equiv & J-v_\phi, \\
\label{53}
J_{2\gamma} & \equiv & J_t,
\end{eqnarray}
then, when an orbit moves from one region of phase space to another one, the variations of $J_2$ are essentially due to the same geometrical terms as those derived in Paragraph~\ref{II.2}. Consequently, one may define an adiabatic invariant $\mJ_2$ for $H_2$ by replacing $I$ by $J_2$ in Eq.~(\ref{35}). \

\subsection{Comparisons between the results obtained by using $H$ or $H_2$}

\begin{figure}[!h]
\centerline{\includegraphics[width=12cm]{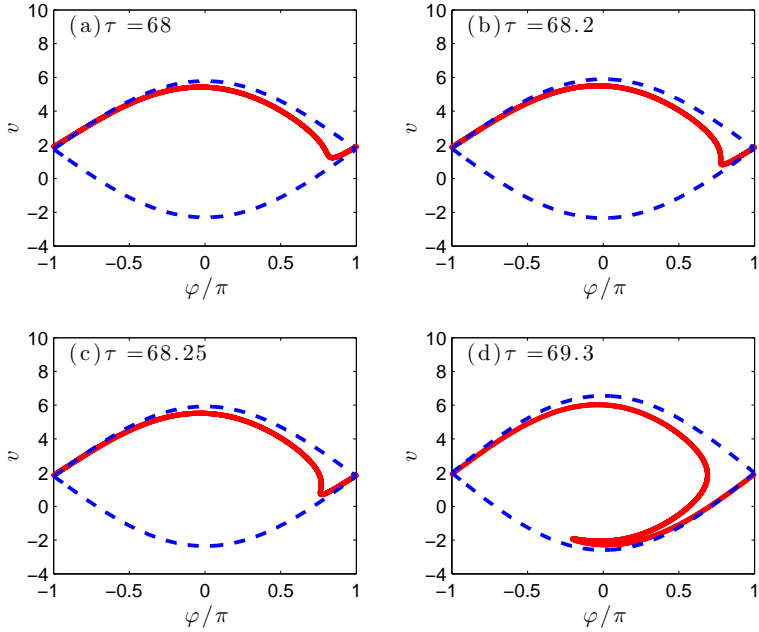}}
\caption{\label{fplus} (Color online) The separatrix (blue dashed line) and the position in phase space of the particles orbits (red lines), as obtained by solving numerically Eqs. (\ref{6}) and (\ref{7}). } 
\end{figure}

In the adiabatic regime, one expects $\dot{v}_\phi \ll \Phi$ (see Appendix~\ref{A} for details), so that $I$ and $J_2$ are close to each other in each region, as should be. Indeed, the adiabatic predictions regarding the transition from one region to another should not vary much depending on the choice made for the Hamiltonian. Nevertheless, they are not exactly the same, which motivates the numerical testing of their relative accuracy. For a wave that keeps growing, we compare the theoretical predictions for the trapping time with numerical results. Fig.~\ref{fplus} shows one example of such a numerical test. For this example, $\Phi=\Phi_0 \left[e^{\gamma_1 \tau}-1\right]$ and $v_\phi=v_{\phi_0}\left[e^{\gamma_2 \tau}-1Ê\right]$ with $\Phi_0=10^{-5}$, $v_{\phi_0}=1.94\times10^{-3}$, $\gamma_1=0.19$ and $\gamma_2=0.1$. As for the particles, we assume that they all have the same initial velocity, $v_0=3.914$. With these parameters, we expect $\dot{v}_\phi/\Phi \approx 5\times10^{-2}$ when trapping occurs, so that adiabatic predictions should be relevant, while not extremely accurate. Using Hamiltonian $H$, we find that the particles should be trapped at $\tau \equiv \tau^{*} \approx 67$, while using $H_2$ we find $\tau^{*}Ê\approx 66.3$. As expected, the predictions for the trapping time are close to each other, but not exactly the same.

Numerically, we solve Eqs.~(\ref{6}) and~(\ref{7}) with 8192 initial normalized positions, $\varphi$, uniformly distributed between $-\pi$ and $\pi$, and the same initial velocity, $v_0=3.914$.  We estimate that trapping starts when the curve $\varphi(v)$ is no longer single-valued, which happens for the first time when $\tau \approx 68.25$, as may be seen in Fig.~\ref{fplus}. Therefore, from this example, we conclude that using $H$ leads to more accurate predictions as regards trapping than using $H_2$. We came to the same conclusion for all the examples we numerically investigated when $\dot{v}_\phi/\Phi \ll 1$. %This is not surprising because, in order to derive the time evolution of $I_2$, we had to approximate $dI_2/dt$ by $\langle dI_2/dt \rangle$ and to approximate $\langle dI_2/dt \rangle$ by the right-hand side of Eq.~(\ref{48}). Hence, we made use of more approximate results than when assuming $I \approx Const.$ 
Moreover, formulas derived from  $H$ are easier to manipulate since they provide an explicit expression for the action, while when using $H_2$ one has to calculate the integral $J$ defined by Eq.~(\ref{14}). Consequently, in Sections~\ref{III}~to~\ref{V}, we will only use results obtained from $H$.\\

Now, as discussed in Appendix~\ref{B}, $H_2$ is best suited to study the dynamics if $\dot{v}_\phi > \Phi$ when the conditions for separatrix crossing, Eq.~(\ref{49}) or Eq.~(\ref{50}), are fulfilled. In particular, no trapping may occur when $\dot{v}_\phi > \Phi$ which is easier to derive by using $H_2$ rather than $H$. However, as shown in Appendix~\ref{A}, the dynamics is not adiabatic when $\dot{v}_\phi>\Phi$. Consequently, this situation is disregarded in the main text and is only considered in Appendix~\ref{B} for the sake of completeness.

\section{Transition probabilities, and numerical results}
\label{III}
\subsection{Derivation of the transition probabilities}
\label{III.1}
The full derivation of the transition probabilities, which we denote by $p^{(\xi\rightarrow\zeta)}$ for an orbit moving from region ($\xi$) to region ($\zeta$), can be found in Refs.~\cite{cary,neishtadt}. However,  it constitutes only one small part of these papers, which are  mathematically involved. Therefore, we found it useful to recall here the main steps of the derivation.  \\ 

Before this, let us first consider situations when the transition probabilities are obvious. From Eq.~(\ref{49}), we know that an orbit leaves region ($\alpha$) when $I=v_{tr}+v_\phi$ and $\dot{v}_{tr}+\dot{v}_\phi >0$. If $\dot{v}_{tr}-\dot{v}_\phi >0$, we know from Eq.~(\ref{50}) that the orbit cannot remain in region~($\beta$). Therefore, it may only move to region $\gamma$, and $p^{(\alpha\rightarrow\gamma)}=1$. By contrast, from Eq.~(\ref{51b}), if $\dot{v}_{tr}<0$, the orbit cannot stay in region ($\gamma$), so that $p^{(\alpha\rightarrow\beta)}=1$.

Using the same kind of reasoning, one easily finds that, when  $I={v}_{tr}-{v}_\phi$ and $\dot{v}_{tr}-\dot{v}_\phi>0$, $p^{(\beta\rightarrow\gamma)}=1$ if $\dot{v}_{tr}+\dot{v}_\phi >0$ while $p^{(\beta\rightarrow\alpha)}=1$ if $\dot{v}_{tr}<0$.

Similarly, when $I={v}_{tr}$ and $\dot{v}_{tr}<0$, $p^{(\gamma\rightarrow\alpha)}=1$ if $\dot{v}_{tr}-\dot{v}_\phi >0$ and $p^{(\gamma\rightarrow\beta)}=1$ if $\dot{v}_{tr}+\dot{v}_\phi >0$. \\

Now, when none of the situations discussed above occurs, an orbit leaving one given region may end up in either of the two other ones with probabilities that we now calculate. To do so, we use Hamilton equations expressing the $\varphi$-evolution of the dynamics, instead of its time evolution. Namely, we invert Eq.~(\ref{H}) to express $v$ as a function of $\varphi$ and $\tau$ so that, using the chain rule, Hamilton equations read~\cite{cary}, 
\begin{eqnarray}
\label{59}
\frac{d \tau}{dÊ\varphi}Ê&=& \left. \frac{\partial v}{\partial H}\right\vert_{\tau}, \\
\label{60}
\frac{d H}{dÊ\varphi}Ê&=&-\left. \frac{\partial v}{\partial \tau}\right \vert_{H}.
\end{eqnarray}
Let us now introduce 
\begin{equation}
\label{61}
h \equiv H-\Phi.
\end{equation}
From the results of Section~\ref{II}, it is quite clear that $h>0$ in regions $(\alpha)$ and $(\beta)$ and $h<0$ in region $(\gamma)$. Moreover, since $h-H$ only depends on time, $h$ is a valid Hamiltonian to study the particles dynamics. 

During a separatrix crossing from region $(\alpha)$, a particle first moves from $\varphi \approx -\pi$ to $\varphi \approx \pi$ along an orbit that is very close to the upper branch of the separatrix, and then from $\varphi \approx \pi$ to $\varphi \approx -\pi$ along an orbit that is very close to the lower branch of the separatrix. According to Eq.~(\ref{60}), during the motion from $\varphi \approx -\pi$ to $\varphi \approx \pi$, $h$ has varied by,
\begin{equation}
\label{63}
\delta h_1= -\partial_t \int_{-\pi}^{\pi} vd\varphi.
\end{equation}
Assuming $h=0$ (which is the value on the separatrix), one easily finds from the results of Section~\ref{II}~that,
\begin{equation}
\label{64}
\delta h_1 \approx -(\dot{\mA}+\dot{v}_\phi).
\end{equation}
From Eq.~(\ref{64}), the orbits leaving region $(\alpha)$ are such that $0<h<\dot{\mA}+\dot{v}_\phi$ [which requires $\dot{\mA}+\dot{v}_\phi>0$ in agreement with Eq.~(\ref{49})], so that $h$ is indeed small. Similarly, when $\varphi$ varies from $\varphi \approx \pi$ to $\varphi \approx -\pi$, $h$ changes by,
\begin{equation}
\label{65}
\delta h_2 \approx -(\dot{\mA}-\dot{v}_\phi),
\end{equation}
so that, during the whole round trip, the value of $h$ has been shifted by,
\begin{equation}
\label{66}
\delta h \approx -2\dot{\mA}.
\end{equation}
After this round trip, the orbit ends up in region $(\beta)$ if and only if $h+\delta h>0$, i.e., if $h>2\dot{\mA}$ [which requires $\dot{\mA}-\dot{v}_\phi<0$ in agreement with Eq.~(\ref{50})]. Hence, if one considers a set of orbits, with uniform distribution in $h$, the fraction of such orbits that end up in region $(\beta)$ after leaving region $(\alpha)$ is just $ (\dot{v}_\phi-\dot{\mA})/(\dot{\mA}+\dot{v}_\phi)$. Therefore, the probability to move from region $(\alpha)$ to region $(\beta)$ is,
\begin{equation}
\label{67}
p^{(\alpha \rightarrow \beta)}=\min\left[\max\left(\frac{\dot{v}_\phi-\dot{\mA}}{\dot{v}_\phi+\dot{\mA}},0 \right),1\right],
\end{equation}
so that the probability to move from region $(\alpha)$ to region $(\gamma)$ is,
\begin{equation}
\label{68}
p^{(\alpha \rightarrow \gamma)}=\min\left[\max\left(\frac{2\dot{v}_{tr}}{\dot{v}_\phi+\dot{v}_{tr}},0 \right),1\right].
\end{equation}
Similarly, one easily finds,
\begin{eqnarray}
\label{69}
p^{(\beta \rightarrow \alpha)}&=&\min\left[\max\left(\frac{\dot{v}_{tr}+\dot{v}_\phi}{\dot{v}_\phi-\dot{v}_{tr}},0 \right),1\right],\\
\label{70}
p^{(\beta \rightarrow \gamma)}&=&\min\left[\max\left(\frac{2\dot{v}_{tr}}{\dot{v}_{tr}-\dot{v}_\phi},0 \right),1\right],\\
\label{71}
p^{(\gamma \rightarrow \alpha)}&=&\min\left[\max\left(\frac{\dot{v}_{tr}+\dot{v}_\phi}{2\dot{v}_{tr}},0 \right),1\right],\\
\label{72}
p^{(\gamma \rightarrow \beta)}&=&\min\left[\max\left(\frac{\dot{v}_{tr}-\dot{v}_\phi}{2\dot{v}_{tr}},0 \right),1\right].
\end{eqnarray}
Eqs.~(\ref{67})-(\ref{72}) hold provided that the distribution in $h$ is uniform, a hypothesis that we now justify. Since the action is nearly conserved, if the initial distributions in action and in its canonically conjugated variable, the angle $\theta$, are uniform, they remain so just before trapping. The same is true if the initial distribution in action is smooth enough, because the range in action corresponding to the range in $h$ involved in separatrix crossing is very narrow. Then, since the change of variables $(I,\theta) \rightarrow (h,\tau)$ is canonical, the distribution in $(h,\tau)$ is also nearly uniform. Moreover, the time $\tau$ at which separatrix crossing occurs depends very little on $h$ (see Ref.~\cite{cary} for details), so that the distribution in $h$ is indeed nearly uniform.

\subsection{Non conservation of the action distribution function}
\label{III.3}
In order to clearly show that the action distribution function is not conserved, and to check the adiabatic predictions, we numerically simulate the motion of particles acted upon by an electrostatic wave such that,
\begin{eqnarray}
\label{85}
\Phi &=&1-\cos(\gamma_1 \tau) \\
v_\phi &=& \gamma_2 \tau,
\end{eqnarray}
with $\gamma_1=10^{-4}$ and $\gamma_2=10^{-5}$. The particles all have the same initial velocity, $v_0=1$, and the 32768 initial values we choose for $\varphi$ are uniformly distributed between $-\pi$ and $+\pi$. Therefore, the normalized initial distribution in action is,
\begin{equation}
\label{86}
f(I,\tau=0)=\delta(I-1),
\end{equation}
where $\delta(I)$ is the Dirac distribution. Then, by comparing the initial distribution shown in Fig.~\ref{f5}(a) with those obtained numerically and plotted in Figs.~\ref{f6}(d),~\ref{f7}(d),~and \ref{f8}(d), it is quite clear that the action distribution function is not conserved. 

Let us now comment the numerical results in more detail, and let us compare them with the theoretical predictions of Section~\ref{II}. 

\subsubsection{First trapping}
\begin{figure}[!h]
\centerline{\includegraphics[width=15cm]{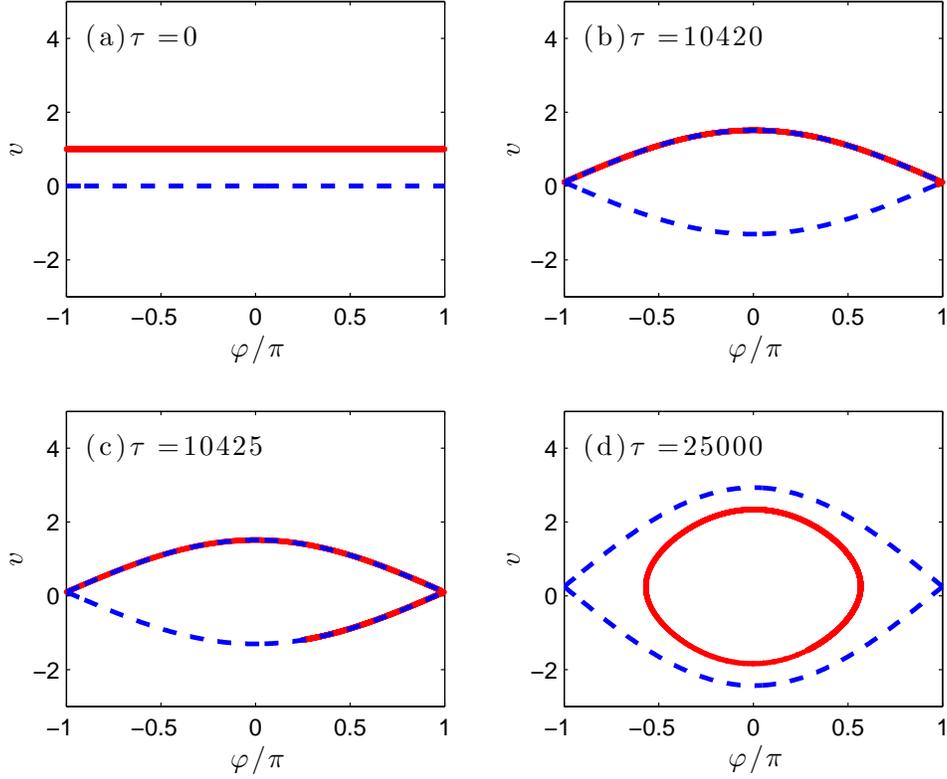}}
\caption{\label{f5} (Color online) The separatrix (blue dashed line) and the positions in phase space of the particles orbits (red curves), as obtained numerically. Panels(b) and (c) show that separatrix crossing occurs at $\tau \approx 10420$ while panel (d) shows that it leads to the trapping of all particles.}
\end{figure}

Since all the particles are initially in region $(\alpha)$ of phase space, from Eq.~(\ref{49}) a first separatrix crossing should occur at $\tau=\tau_1 \approx  10415$ when $v_0-v_\phi=v_{tr}$. Since at this time, $\dot{v}_{tr}Ê\approx 7.8\times 10^{-5}Ê> \dot{v}_\phi$, separatrix crossing should lead to the trapping of all particles. These predictions are in excellent agreement  with the numerical results, as shown in Fig.~\ref{f5}. Moreover, since at time $\tau=\tau_1$, $v_\phi \approx 0.1$, one concludes from Eq.~(\ref{30}) that, after trapping, the new action is $I_{\gamma}Ê\approx 0.9$.  

Now, it is noteworthy that, if we reverse the velocities of the trapped particles, their action does not change. Moreover, if we reverse time after the particles have been trapped and if we assume that their motion is adiabatic, we would conclude from the results of Section~\ref{II} that, after  detrapping, the initial distribution function would not be recovered. Indeed, when detrapping would occur, $\vert \dot{v}_\phi \vert < \vert \dot{v}_{tr}Ê\vert$ so that the particles would be distributed into two distinct sets, in a fashion similar to what is illustrated in Fig.~\ref{f6}(d). Hence, by making use of the adiabatic approximation, one looses the microreversibility of the dynamics. In practice, the particles indeed seem to evolve in an irreversible  way since their distribution function becomes more and more complex as time goes by [see Fig.~\ref{f8}(d)]. 

\subsubsection{First detrapping}
\begin{figure}[!h]
\centerline{\includegraphics[width=15cm]{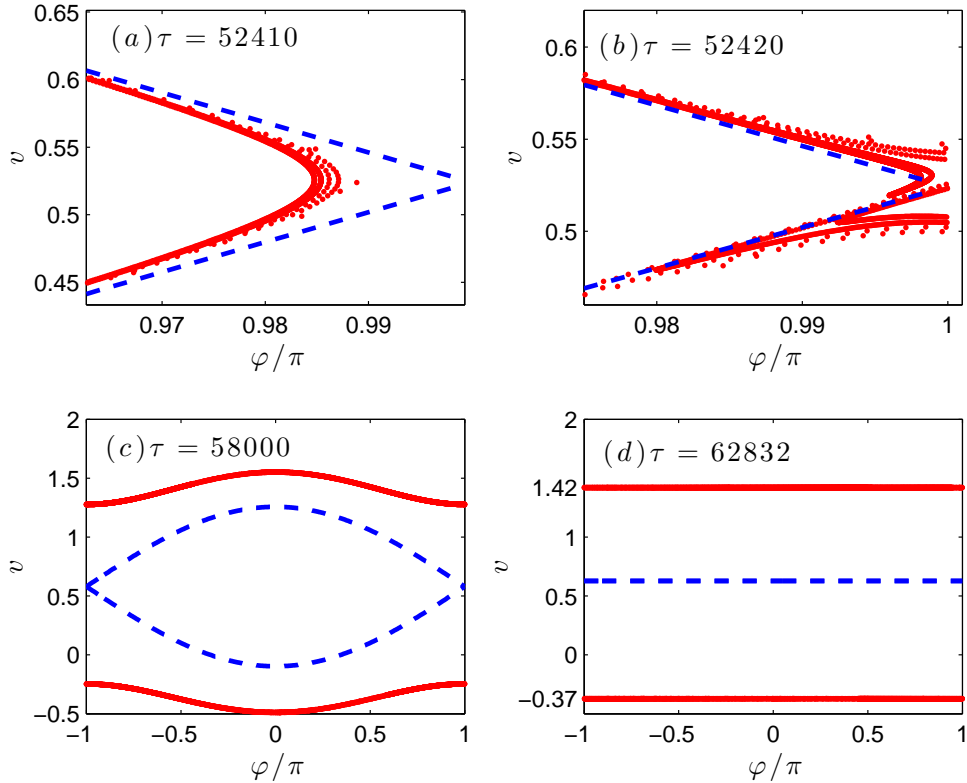}}
\caption{\label{f6} (Color online) The separatrix (blue dashed line) and the positions in phase space of the particles orbits (red curves), as obtained numerically. In panels (a) and (b), we zoomed over positions close to $\varphi = \pi$ to show that, at time $\tau=52410$, all the particles are still trapped while some are detrapped at $\tau=52420$, leading to the splitting of the distribution in action. This splitting is more obvious in panel (c), while in panel (d) one may  clearly see that the particles action is either, $I_\alpha \approx 1.42$ or $I_\beta \approx 0.37$. }
\end{figure}

From the evolution of the wave given by Eqs.~(\ref{85})~and~(\ref{86}), and using Eq.~(\ref{51b}), one easily finds that the particles should be detrapped at $\tau_2=2\pi/\gamma_1-\tau_1 \approx 52417$. Moreover since, at this time, $\vert\dot{v}_{tr}\vertÊ\approx 7.8\times 10^{-5}Ê> \dot{v}_\phi$, the particles should be split between the regions $(\alpha)$ and $(\beta)$ of phase space. Again, this is in excellent agreement with the numerical results illustrated in Fig.~\ref{f6}. Hence, after time $\tau_2$, there are two classes of passing particles, one in region $(\alpha)$ and one in region $(\beta)$. Within each class, all the particles nearly have the same action, given respectively by Eqs.~(\ref{33})~and~(\ref{34}) with $v_\phi \approx 0.52$,
\begin{eqnarray}
\label{87}
I_{\alpha}Ê&=& I_{\gamma}+v_\phi\approx 1.42003, \\
\label{88}
I_{\beta}Ê&=& I_{\gamma}-v_\phi\approx 0.3717.
\end{eqnarray}
These predictions may be easily checked numerically since, from the very definition Eq.~(\ref{10}) of $I$ for passing particles, when $\Phi=0$, $I_\alpha=\langle v \rangle$ and $I_\beta=-\langle v \rangle$. Numerically, we find respectively in region $\alpha$ and $\beta$,
\begin{eqnarray}
\label{89}
\langle v_\alpha \rangle_{num}Ê\approx 1.41996, \\
-\langle v_\beta \rangle_{num}Ê\approx 0.3715,
\end{eqnarray}
where $\langle . \rangle_{num}$ actually stands for a statistical averaging over all the particles in a given region of phase space at time $\tau=2\pi/\gamma_1$ when $\Phi=0$. Hence, the agreement between the numerical results and the theoretical predictions is excellent. 

\subsubsection{Second and third trappings}
\begin{figure}[!t]
\centerline{\includegraphics[width=15cm]{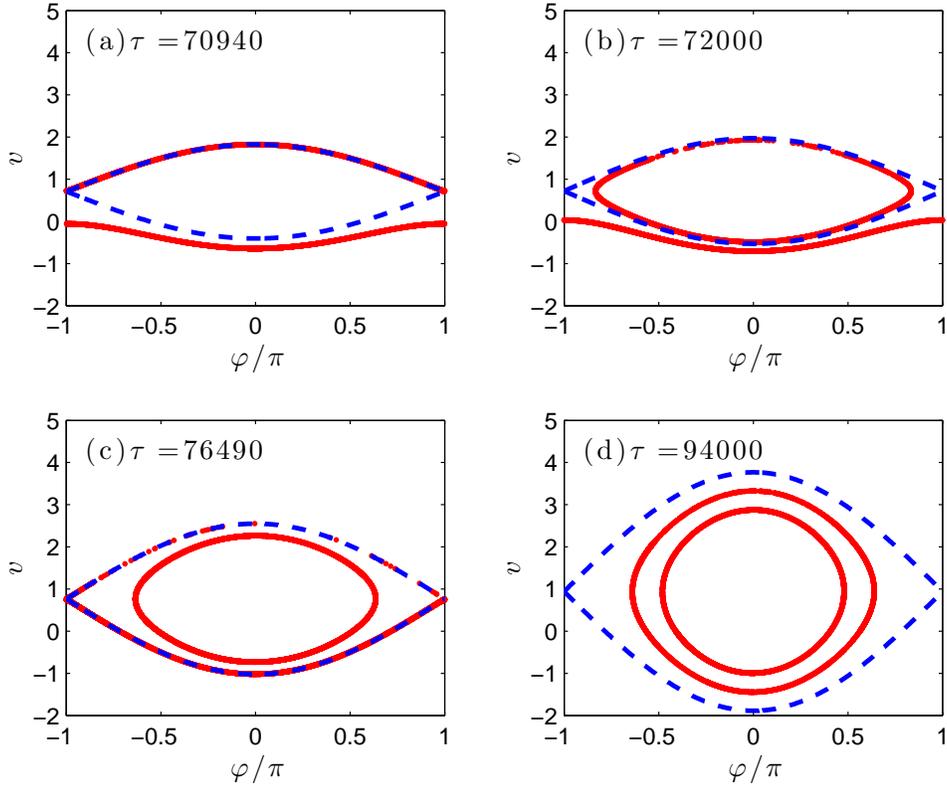}}
\caption{\label{f7} (Color online) The separatrix (blue dashed line) and the positions in phase space of the particles orbits (red curves), as obtained  numerically. Panel (a) shows that a first separatrix crossing occurs at $\tauÊ\approx 70940$ while panel (b) shows that it leads to the trapping of all the particles which were in region~$(\alpha)$ of phase space. Moreover, panel (c) shows that the particles orbits in region $(\beta)$  cross the separatrix when $\tau \approx 76940$, which leads to essentially two very distinct populations of trapped particles, as illustrated in panel (d). }
\end{figure}

Using again Eqs.~(\ref{49})~and~(\ref{50}), one finds that the particles in region $\alpha$ should be the first to be trapped again, at time $\tau_3 \approx  70945$ when $I_\alpha-v_\phi=v_{tr}$. At this time, $\dot{v}_{tr}Ê\approx 8.3\times 10^{-5} >\dot{v}_\phi$, so that all the particles should be trapped. As may be seen in Figs.~\ref{f7}(a) and (b), this is in excellent agreement with the numerical results. Since at time $\tau_3$, $v_\phi \approx 0.71$, the action of the first trapped particles is,
\begin{equation}
\label{91}
I_{\gamma_1}=I_\alpha-v_\phiÊ\approx 0.71.
\end{equation}
Similarly, Êfrom the results of Section~\ref{II} one would conclude that all the particles in region $\beta$ should be trapped at time $\tau =\tau_4 \approx 76495$, when $I_\beta+v_\phi=v_{tr}$. At this time, $v_\phi \approx 0.76$, so that the action of the trapped particles is,
\begin{equation}
\label{92}
I_{\gamma_2}=I_\beta+v_\phiÊ\approx 1.14.
\end{equation}
Hence, after time $\tau_4$ there should be two distinct populations of trapped particles each having its own action even though, initially, all the particles had exactly the same action. Again, the theoretical predictions are in excellent agreement with the numerical results plotted in Figs~\ref{f7}(c) and (d).  \\

\subsubsection{Second and third detrappings}
\begin{figure}[!h]
\centerline{\includegraphics[width=15cm]{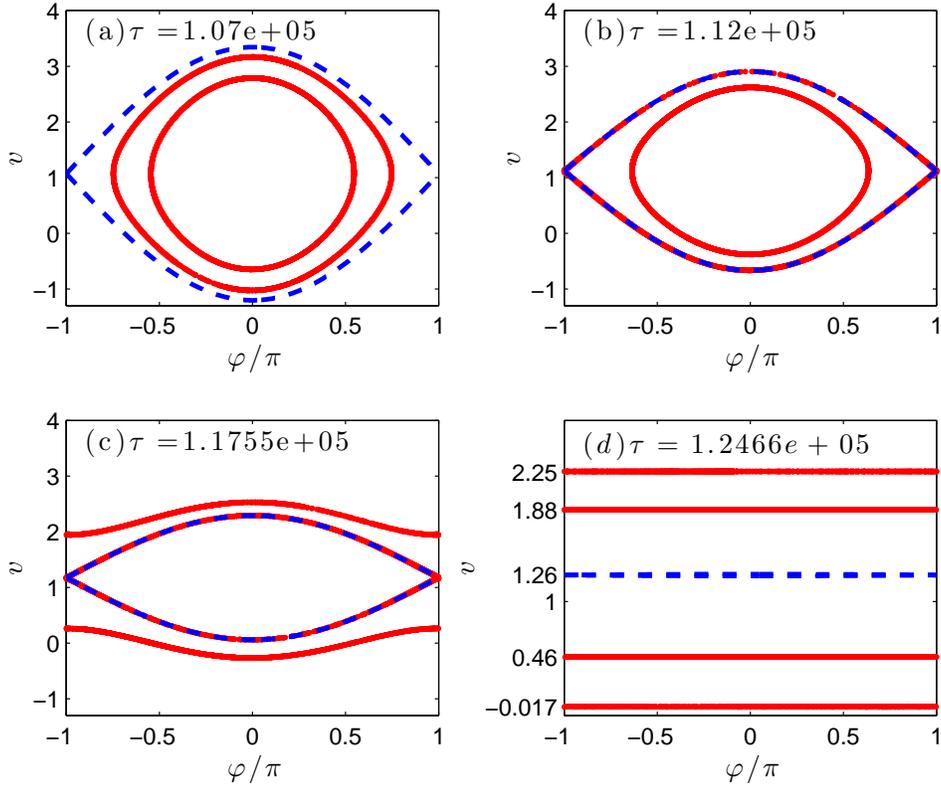}}
\caption{\label{f8} (Color online) The separatrix (blue dashed line) and the positions in phase space of the particles orbits (red curves), as obtained  numerically. Panel (b) and (c) show that the two sets of trapped particles are detrapped respectively at $\tau \approx 1.12\times 10^{5}$ and $\tau \approx 1.1755\times 10^{5}$. This leads to essentially four distinct populations of passing particles with actions, $I_{\beta_1}Ê\approx 0.017$, $I_{\beta_2}Ê\approx -0.46$, $I_{\alpha_2}Ê\approx 1.88$ and $I_{\alpha_1}Ê\approx 2.25$, as shown in panel (d).}
\end{figure}
Using again Eq.~(\ref{51b}), one would find that particles with action $I_{\gamma_2}$ should be the first to be detrapped  when $\tau=\tau_5=6\pi/\gamma_1-\tau_4 \approx 1.12\times 10^{5}$, leading to two sets of passing particles, in excellent agreement with the numerical results shown in Fig.~\ref{f8}. When this occurs, $v_\phi \approx 1.12$, so that the actions in region $\alpha$ and $\beta$ should be
\begin{eqnarray}
\label{94}
I_{\alpha_1}Ê=I_{\gamma_2}+v_\phi \approx 2.25, \\
\label{95}
I_{\beta_1}Ê=I_{\gamma_2}-v_\phi \approx 0.017,
\end{eqnarray}
in excellent agreement with the results shown in Fig.~\ref{f8}(d). 

Similarly, using the results of Section~\ref{II}, one would find that  detrapping would occur again when $I_{\gamma_1}=v_{tr}$ at $\tau=\tau_6 =6\pi/\gamma_1-\tau_3\approx 1.1755\times 10^{5}$ leading to two sets of passing particles with actions, respectively in regions $(\alpha)$ and $(\beta)$,
\begin{eqnarray}
\label{96}
I_{\alpha_2}Ê=I_{\gamma_1}+v_\phi \approx 1.88, \\
\label{97}
I_{\beta_2}Ê=I_{\gamma_2}-v_\phi \approx -0.46.
\end{eqnarray}
This is, again, in excellent agreement with the numerical results shown in Fig.~\ref{f8}(d).

\subsection{Numerical check of the theoretical predictions for transition probabilities}
\label{III.4}
In Paragraph \ref{III.3}, we have just shown an excellent agreement between the numerical results and the adiabatic predictions of Section~\ref{II} as regards the times when separatrix crossings occur. However, we also noted in Paragraph~\ref{III.1}~Êthat the transitions probabilities given by Eqs.~(\ref{67})-(\ref{72}) were only valid if the distribution in action was smooth enough, a condition that was clearly not fulfilled in Paragraph~\ref{III.3} since we chose $f(I,\tau=0)=\delta(I-1)$. Let us now test how accurate Eqs.~(\ref{67})-(\ref{72}) are for the latter choice of $f(I,\tau=0)$, and how the numerical results may be made closer to the theoretical ones by choosing a smoother initial distribution in action. \\

For the first detrapping occurring at $\tau=\tau_2 \approx 52415$ [see Figs.~\ref{f6}(a)~and~(b)], $\dot{v}_{tr}Ê\approx -7.8\times 10^{-5}$ and $\dot{v}_\phi=10^{-5}$. Then, from Eqs.~(\ref{71})~and~(\ref{72}) one finds,
\begin{eqnarray}
\label{100}
p^{(\gamma \rightarrow \alpha)}_{adia}(\tau_2)& \approx & 43.5\%, \\
\label{101}
p^{(\gamma \rightarrow \beta)}_{adia}(\tau_2)& \approx & 56.5\%.
\end{eqnarray}
Numerically, after time $\tau_2$ (and before time $\tau_3$), we find that 19146 particles orbits are in region~$(\alpha)$ and 13622 are in region~$(\beta)$. Hence,  when $\tau \approx \tau_2$, about 58\% of all particles orbits moved from region $(\gamma)$ to region $(\alpha)$, and about 42\% of them went to region $(\beta)$ so that, numerically, one finds,
\begin{eqnarray}
\label{102}
p^{(\gamma \rightarrow \alpha)}_{num}(\tau_2)& \approx & 58\%, \\
\label{103}
p^{(\gamma \rightarrow \beta)}_{num}(\tau_2)& \approx & 42\%.
\end{eqnarray}
Therefore, the agreement between the numerical results and the theoretical predictions are not good. In particular, we numerically find that most orbits move from region $(\gamma$) to region $(\alpha)$, while the highest transition probability was predicted to be $p^{(\gamma \rightarrow \beta)}$.

Similarly, for the detrapping occurring at times $\tau_5 \approx 1.2\times 10^{5}$ and $\tau_6 \approx 1.1755\times 10^{5}$ (see Fig.~\ref{f8}), the adiabatic predictions are,
\begin{eqnarray}
\label{104}
p^{(\gamma \rightarrow \alpha)}_{adia}(\tau_5)& \approx & 43\%, \\
\label{105}
p^{(\gamma \rightarrow \beta)}_{adia}(\tau_5)& \approx & 57\%, \\
\label{106}
p^{(\gamma \rightarrow \alpha)}_{adia}(\tau_6)& \approx & 44\%, \\
\label{107}
p^{(\gamma \rightarrow \beta)}_{adia}(\tau_6)& \approx & 56\%.
\end{eqnarray}
Numerically we find that, just after time $\tau_5$,  5479 particles orbits are in region~$(\alpha)$ and 8143 in region $(\beta)$, while just after time $\tau_6$, 10360 additional particles orbits have moved to region~$(\alpha)$ and 8786 to region $(\beta)$ so that,
\begin{eqnarray}
\label{108}
p^{(\gamma \rightarrow \alpha)}_{num}(\tau_5)& \approx & 40\%, \\
\label{109}
p^{(\gamma \rightarrow \beta)}_{num}(\tau_5)& \approx & 60\%, \\
\label{110}
p^{(\gamma \rightarrow \alpha)}_{num}(\tau_6)& \approx & 54\%, \\
\label{111}
p^{(\gamma \rightarrow \beta)}_{num}(\tau_6)& \approx & 46\%.
\end{eqnarray}
Again, we find the adiabatic transition probabilities are not very reliable. This is because the distribution in action is not smooth.\\ %As proven in Appendix~\ref{D}, for an initial Dirac distribution in action, the distribution in $h$ is not uniform, it is the sum of two Dirac distributions, 
%\begin{equation}
%\label{n11}
%f(h)=\frac{T(h_{\min})\delta(h-h_{\min})+T(h_{\max})\delta(h-h_{\max})}{T(h_{\min})+T(h_{\max})},
%\end{equation}
%where $h_{\min}$ and $h_{\max}$ are the minimum and maximum values of $h$. As for $T(h)$ it is the period of the frozen orbit associated with $h$. \\

In order to get a better agreement between the numerical results and the theoretical predictions, we performed another simulation very similar to the one discussed  in Paragraph~{III.3}~except that, now, the initial distribution in action is,
\begin{equation}
\label{112}
f(I,\tau=0)= Y(v_0-\delta v)-Y(v_0+\delta v),
\end{equation}
where $Y(v)$ is  the Heaviside distribution, $v_0=1$ and $\delta v=10^{-3}$. The distribution in Eq.~(\ref{112}) is actually not smooth, nevertheless it is much smoother than a Dirac one. Numerically, in order to simulate such a distribution, we just randomly choose the initial particles positions and velocities in the domain $[-\pi,\pi]\times [v_0-\delta v,v_0+\delta v]$. The phase portraits we obtain are very similar to those illustrated in Figs.~\ref{f5}-\ref{f8} and, therefore, they will not be reproduced here. Moreover, as regards the transition probabilities we find that, just after time $\tau_2$, 14251 particles orbits are in region $(\alpha)$ and 18517 are in region $(\beta)$. Just after time $\tau_5$, 8215 particles orbits are in region $(\alpha)$ and 10302 are in region $(\beta)$, and just after time $\tau_6$, 6384 additional ones are in region $(\alpha)$ and 7867 in region $(\beta)$. This leads to the following results, 
\begin{eqnarray}
\label{112}
p^{(\gamma \rightarrow \alpha)}_{num}(\tau_2)& \approx & 43.5\%, \\
\label{113}
p^{(\gamma \rightarrow \beta)}_{num}(\tau_2)& \approx & 56.5\%, \\
\label{114}
p^{(\gamma \rightarrow \alpha)}_{num}(\tau_5)& \approx & 44\%, \\
\label{115}
p^{(\gamma \rightarrow \beta)}_{num}(\tau_5)& \approx & 56\%, \\
\label{116}
p^{(\gamma \rightarrow \alpha)}_{num}(\tau_6)& \approx & 45\%, \\
\label{117}
p^{(\gamma \rightarrow \beta)}_{num}(\tau_6)& \approx & 55\%.
\end{eqnarray}
Hence, the agreement with the adiabatic predictions, Eqs~(\ref{100}), (\ref{101}), (\ref{104})-(\ref{107}) is, now, very good. It remains so provided that $\delta v \agt 2\times 10^{-4}$. This makes sense, because the action distribution function has to remain nearly constant over a range, $\Delta I > \max(\gamma_1,\gamma_2)$, for the adiabatic results to be valid. More generally, and coming back to physical units, let us denote by $I_T$ the typical range of variation in the action distribution function about the wave phase velocity when separatrix crossing occurs, by $\Gamma$  the typical rate of variation of the wave amplitude and phase velocity (or the largest of these two rates), and by $k$ the wave number. Then, the transition probabilities are accurate provided that $I_T > \Gamma/k$ (see Ref.~\cite{benisti16} for an application of this result). The latter condition is equivalent to Eq.~(\ref{n3}) derived in Appendix~\ref{A0}.
 
\section{The adiabatic distribution function}
\label{IV}
Let us now derive our main result regarding the time evolution of the action distribution function. We only restrict to smooth enough distributions [as defined in Paragraph \ref{III.4} or, equivalently, by Eq.~(\ref{n3})] so that the adiabatic transition probabilities are expected to be accurate. Moreover,  instead of deriving the distribution function at any time, which would lead to very complicated formulas, we just explain how it evolves each time a separatrix crossing occurs (while it is known to remain constant when there is no crossing). 

Let us consider a transition from region $(\xi)$ to region $(\zeta)$ (where $\xi$ and $\zeta$ are either $\alpha$, $\beta$ or $\gamma$), and let us denote by $f^{<}(I)$ the action distribution function just before the transition, and by $f^{>}(I)$ the distribution just after. Let us, moreover, denote by $\delta I$ the geometrical jump in action entailed by separatrix crossing (and derived in Paragraph~\ref{II.2}). Then,
\begin{equation}
\label{401}
f^{>}(I)=f^{<}(I-\delta I) p^{(\xi\rightarrow \zeta)} \left \vert 1-\frac{d\deltaÊI}{dI}\right\vert.
\end{equation}
Let us now clarify how $\delta I$ depends on $I$. From the results of Section~\ref{II}, we know that $I=v_{tr}+\delta_{\xi}Êv_{\phi}$, where $v_{tr}$ and $v_\phi$ are calculated at the time when trapping occurs, and $\delta_\alpha=1$, $\delta_\beta=-1$, $\delta_\gamma=0$. Moreover, we also know from Section~\ref{II}, that $\delta I=\mu_{\xiÊ\rightarrow \zeta} v_\phi$, where $\mu_{\xiÊ\rightarrow \zeta} \in \{-2,-1,+1,+1Ê\}$. Therefore, we conclude that

\begin{eqnarray}
\nonumber
\frac{d\deltaÊI}{dI}& =& \frac{\mu_{\xiÊ\rightarrow \zeta} dv_{\phi}}{dv_{tr}+\delta_{\xi}Êdv_{\phi}}\\
\label{402}
&=&\frac{\mu_{\xiÊ\rightarrow \zeta} dv_{\phi}/dv_{tr}}{1+\delta_{\xi}Êdv_{\phi}/dv_{tr}},
\end{eqnarray}
where we have denoted,
\begin{equation}
\label{403}
\frac{dv_\phi}{dv_{tr}} \equiv \frac{\dot{v}_\phi}{\dot{v}_{tr}}.
\end{equation}
Using Eqs.~(\ref{401})-(\ref{403}), together with Eqs.~(\ref{67})-(\ref{72}) for the transition probabilities, one straightforwardly  finds the following results. \\

Transition $\alpha \rightarrow \beta$, $I=v_{tr}-v_\phi$, $\dot{v}_{tr}+\dot{v}_\phi >0$ and $\dot{v}_\phi-\dot{v}_{tr} >0$, 
\begin{equation}
\label{404}
f^{>}_{\beta}(I)=f_{\alpha}^{<}(I+2v_\phi) \min\left[\frac{\dvdv-1}{\dvdv+1},1\right]  \frac{\dvdv+1}{\dvdv-1}. 
\end{equation}
\newline 

Transition $\alpha \rightarrow \gamma$, $I=v_{tr}$, $\dot{v}_{tr}+\dot{v}_\phi >0$ and $\dot{v}_{tr}>0$,
\begin{equation}
\label{405}
f_{\gamma}^{>}(I)=f_{\alpha}^{<}(I+v_\phi) \min\left[\frac{2}{\dvdv+1},1\right]  (1+\dvdv).
\end{equation}
\newline 

Transition $\beta \rightarrow \alpha$, $I=v_{tr}+v_\phi$, $\dot{v}_\phi-\dot{v_{tr}} <0$ and $\dot{v}_\phi +\dot{v}_{tr}<0$,
\begin{equation}
\label{406}
f_{\alpha}^{>}(I)=f_{\beta}^{<}(I-2v_\phi) \min\left[\frac{\dvdv+1}{\dvdv-1},1\right]  \frac{\dvdv-1}{\dvdv+1}. 
\end{equation}
\newline 

Transition $\beta \rightarrow \gamma$, $I=v_{tr}$, $\dot{v}_{tr}-\dot{v}_\phi >0$ and $\dot{v}_{tr}>0$,
\begin{equation}
\label{407}
f_{\gamma}^{>}(I)=f_{\beta}^{<}(I-v_\phi) \min\left[\frac{2}{1-\dvdv},1\right]  (1-\dvdv).
\end{equation}
\newline 

Transition $\gamma \rightarrow \alpha$, $I=v_{tr}+v_{\phi}$, $\dot{v}_{tr}<0$ and $\dot{v}_{tr}+\dot{v}_\phi <0$,
\begin{equation}
\label{408}
f_{\alpha}^{>}(I)=f_{\gamma}^{<}(I-v_\phi) \min\left[\frac{1+\dvdv}{2},1\right]  \frac{1}{1+\dvdv}.
\end{equation}
\newline 

Transition $\gamma \rightarrow \beta$, $I=v_{tr}-v_{\phi}$,  $\dot{v}_{tr}<0$ and $\dot{v}_{tr}-\dot{v}_\phi <0$,
\begin{equation}
\label{409}
f_{\beta}^{>}(I)=f_{\gamma}^{<}(I+v_\phi) \min\left[\frac{1-\dvdv}{2},1\right]  \frac{1}{1-\dvdv}.
\end{equation}
\newline 

Transitions $\beta \rightarrow \gamma$ and $\alpha \rightarrow \gamma$ occur simultaneously whenever $I=v_{tr}$, $\dot{v}_{tr}>0$, $\dot{v}_{tr}+\dot{v}_\phi >0$ and $\dot{v}_{tr}-\dot{v}_\phi >0$. Then, 
\begin{eqnarray}
\nonumber
f_{\gamma}^{>}(I)=	&&f_{\alpha}^{<}(I+v_\phi) \min\left[\frac{2}{\dvdv+1},1\right]  (1+\dvdv) \\
\label{410}
&&+f_{\beta}^{<}(I-v_\phi) \min\left[\frac{2}{1-\dvdv},1\right]  (1-\dvdv).
\end{eqnarray}
\newline 

Note that, when deriving the action distribution function, we explicitly indicated the corresponding region in phase space because, with our definition for $I$, different orbits in different regions of phase space might have the same action. Indeed, in region $(\gamma)$, $0 \leq I \leq v_{tr}$, in region $(\alpha)$, $I \geq v_{tr}+v_\phi$ and in region $(\beta)$, $I \geq v_{tr}-v_\phi$. For the sake of definiteness, let us assume that $v_\phi>0$. Then, for any value $I$ such that $0\leq I \leq v_{tr}$, there exists one orbit in region $(\gamma)$ and one orbit in region $(\beta)$. Similarly, for any value $I$ such that $I>v_{tr}+v_\phi$, there exists one orbit in region $(\alpha)$ and one orbit in region $(\beta)$. Consequently, $f(I)$ cannot be defined unambiguously in the whole phase space, but only within each sub-region. 

Note also that, when the transition probabilities are less than unity, the left-hand sides and right-hand sides of Eqs.~(\ref{404})-(\ref{409}) are the same (up to a factor 2 for trapping and detrapping, due to the different definition for the action of trapped particles). Therefore, the transition probabilities have a very simple geometrical interpretation. When they are less than unity, their value is just the relative change in action entailed by separatrix crossing.

\section{Generalization to an inhomogeneous wave}
\label{V}
In order to derive the variations of the action when the wave is inhomogeneous, we essentially use the same method as in Ref.~\cite{benisti16}, which we quickly summarize. In particular, we restrict to the situation when the wave is weakly inhomogeneous, so that the Hamiltonian remains nearly constant within one wave period i.e., when $\varphi$ varies by $2\pi$. 

For trapped particles, the action is still defined by~Eq.~(\ref{13}). Then, $I\equiv I(H,\tau)$ so that $H\equiv H(I,\tau)$, and $dI/d\tau \approx \langle dI/d\tau \rangle=-\partial H/\partial \theta \vert_I=0$. The action of the trapped particles remains nearly constant. Moreover, using the same definition for $m$ as in Paragraph~\ref{II.2} i.e., $mÊ\equiv 2\Phi/(H+\Phi)$, $I$ is still given by Eq.~(\ref{28}), at zero order in the $\varphi$-variations of $\Phi$. 

For the passing particles, the situation is more complex because, now, $H$ is no longer $2\pi$-periodic in $\varphi$. Then, for these particles, the action is necessarily $\varphi$-dependent, and is defined as,
\begin{equation}
\label{501}
I \equiv \frac{1}{2\pi}Ê\int_{\varphi-\pi}^{\varphi+\pi} vd\varphi',
\end{equation}
where the integral is still calculated along a frozen orbit. At zero order in the variations of $\Phi$ and $v_\phi$, $I$ is still given by Eqs.~(\ref{26})~and~(\ref{27}) respectively in regions ($\alpha$) and ($\beta$). However, now, $I\equiv I(H,\varphi,\tau)$ so that $H \equiv H[I,\varphi(\theta,I,\tau),\tau]$. Then, neglecting the variations of $H$ within one wave period, we calculate,
\begin{eqnarray}
\nonumber
\frac{dI}{d\tau}Ê& \approx & \left \langle \frac{dI}{d\tau}\right \rangle \\
\nonumber
&=&- \left.\left\langle \frac{\partial H}{\partial \varphi} \frac{\partial \varphi}{\partial \theta} \right\rangle\right\vert_I \\
\nonumber
& \approx &-\left.\left\langle \frac{\partial H}{\partial \varphi} \right\rangle\right\vert_I \left.\left\langle\frac{\partial \varphi}{\partial \theta} \right\rangle\right\vert_I \\
\nonumber
&=& -\left.\frac{\partial }{\partial \varphi} \left(\frac{[2-m]\Phi}{m}\right)\right\vert_I \\
\label{503}
&=& \left\{1+\frac{2}{m}\left[\frac{E(m)}{K(m)}-1 \right]Ê\right\}\frac{\partial \Phi}{\partial \varphi}+\eta \frac{\pi \sqrt{\Phi}}{\sqrt{m}K(m)}\frac{\partial v_\phi}{\partial \varphi},
\end{eqnarray}
where $\eta=1$ in region ($\alpha$) and $\eta=-1$ in region ($\beta$) [and where we used $\langle \partial \varphi/\partial \theta \rangle\vert_I=1$, and Eqs.~(\ref{26})~and~(\ref{27}) to derive $\partial m/\partial \varphi \vert_I$]. \\

Let us now introduce $\psi$ such that,
\begin{eqnarray}
\nonumber 
\frac{d\psi}{d\tau} &=& \left \langle \frac{d\varphi}{d\tau}\right \rangle \\
\nonumber
& =& \frac{\partial }{\partial I}\left.\left(\frac{[2-m]\Phi}{m}\right)\right\vert_\theta \\
\label{505}
&\approx& \frac{\pi \sqrt{\Phi}}{\sqrt{m}K(m)} ,
\end{eqnarray}
where we used Eqs.~(\ref{26})~and~(\ref{27}), and neglected the small contributions proportional to $\partial_\varphi \Phi$ and $\partial_\varphi v_\phi$. Neglecting the variations of $H$ within one wave period, $\partial H/\partial \varphi \approx \partial H/\partial \psi$, so that  $\psi$ and $I$ may be considered as conjugated variables for $H$. \\

Now, as regards the distribution function of the untrapped particles, which we denote by $f_u$, we assume that it remains nearly constant within one wave period. Then, $f_u(I,\theta,\tau)\approx \langle f_u(I,\theta,\tau) \rangle = f_u(I,\psi,\tau)$ so that, using Eqs.~(\ref{503})~and~(\ref{505}), we straightforwardly find,
\begin{equation}
\label{506}
\frac{\partial f_u}{\partial \tau}+ \frac{\pi \sqrt{\Phi}}{\sqrt{m}K}\frac{\partial f_u}{\partial \psi}+\left\{Ê \left[1+\frac{2}{m}\left(\frac{E}{K}-1 \right)Ê\right]\frac{\partial \Phi}{\partial \varphi}+\eta \frac{\pi \sqrt{\Phi}}{\sqrt{m}K}\frac{\partial v_\phi}{\partial \varphi}\right\}  \frac{\partial f_u}{\partial I}=0.
\end{equation}
Therefore, in addition to the jumps in the action distribution function derived in Section~\ref{IV}, one must also account for the slow and continuous variation of the distribution function of the passing particles, as given by Eq.~(\ref{506}).

\section{Conclusion}
\label{VI}
In conclusion, we investigated the motion of charged particles in a slowly-varying sinusoidal electrostatic wave. When the wave amplitude and phase velocity were space independent, we provided an explicit expression for the adiabatic invariant, $\mI$, which was not the dynamical action. Moreover, $\mI$ was shown to be non-local in the wave variations, thus making the adiabatic dynamics irreversible. 

Using numerical simulations, we provided clear evidences of the non-conservation of the action distribution function. Moreover, we numerically tested the accuracy of adiabatic predictions as regards separatrix crossings and transition probabilities. In particular, we checked that the known expressions for transitions probabilities~\cite{cary,neishtadt} were only relevant to smooth enough action distributions. They must be such that $I_T \agt \Gamma/k$, where $I_T$ is the typical range of variation in the action distribution function about the wave phase velocity when separatrix crossing occurs, $\Gamma$ is the typical rate of variation of the wave amplitude and phase velocity (or the largest of these two rates), and $k$ is the wave number. 

Finally, we explicitly provided the way to derive the time variations of the action distribution function, $f(I)$. When the wave amplitude and phase velocity only depend on time, $f(I)$ experiences a jump each time a separatrix crossing occurs, as given by Eqs.~(\ref{404})-(\ref{410}). When the wave is weakly inhomogeneous, there is no adiabatic invariant for the passing particles. Consequently, in addition to the jumps entailed by separatrix crossings, the action distribution of passing particles slowly varies in space and time following Eq.~(\ref{506}).

Hence, this article provides explicit practical formulas to directly apply adiabatic results. These formulas are used in the companion paper~\cite{companion} to derive the nonlinear frequency shift of an electron plasma wave in a very general fashion, that accounts for plasma inhomogeneity and multidimensional effects.
\begin{acknowledgments}
The authors  would like to thank I.Y. Dodin for useful discussions, and for pointing out Ref.~\cite{breizman}, and J.E. Sevestre for a careful reading of the manuscript.
 \end{acknowledgments} 

\appendix
\section{Slow and fast time scales, and conditions for adiabaticity}
\label{A0}
\newcounter{app}

\setcounter{app}{1}
In this Appendix, we clarify the slow and fast time scales associated with the Hamiltonian $H$ defined by Eq.~(\ref{H}), and we derive the conditions for the neo-adiabatic results of Ref.~\cite{tim,cary,hanna,nei,ten,vas} to apply to $H$. 

As usual, the slow time scale is associated with the variations of the parameters that enter into the definition of $H$. Hence, the slow time scale, which we denote by $T_w$, corresponds to the typical time variation of the normalized wave amplitude, $\Phi$, or of the normalized wave phase velocity, $v_\phi$. Therefore, $T_w$ is associated with the variations of the wave properties.

As for the fast time scale, it is the typical period, $T$, of a frozen orbit away from the separatrix. 

In order to derive when the condition for adiabaticity, $T/T_w \ll 1$, is satisfied, let us first consider an initial Dirac distribution in velocity. Hence we assume that, initially, when the wave amplitude is vanishingly small, all the particles have the same normalized velocity, $v_0$ [so that, from Eq.~(\ref{10}), $I=v_0$]. As shown in Section~\ref{V}, when the particles are untrapped, the period of a frozen orbit is 
\begin{equation}
\label{A1}
T=2\sqrt{m}/K(m)\sqrt{\Phi},
\end{equation}
where $K$ is the elliptic integral of first kind, and where $m$ is defined by Eq.~(\ref{24}). Far away from the separatrix, when $m \ll 1$, $d\varphi/d\tau \approx v_0-v_\phi$, so that $T \approx 2\pi/\vert v_0-v_\phi\vert$. Close to the separatrix, when $m \approx 1$, $T \approx 2\pi/\sqrt{\Phi}$ ($T$ differs from this value by less than 20\% when $0.9<m<0.99$). Now, from Eqs.~(\ref{26})~and~(\ref{27}), $\vert I-v_\phi \vert \approx 4\sqrt{\Phi}/\pi$ when $m \approx 1$, so that $T \approx 8/\vert I-v_\phi\vert$. Since $I=v_0$ we conclude that, whether the particles orbits are close or far away from the separatrix, $T$ is of the order of $2\pi/(v_0-v_\phi)$. This means that the condition,
\begin{equation}
\label{A2}
\varepsilon \equiv 1/(v_0-v_\phi)T_w\ll 1,
\end{equation}
entails the condition for adiabaticity, $T/T_w \ll 1$. 

When the particles are trapped, the period of their frozen orbit is close to 
\begin{equation}
\label{A3}
T \approx 2\pi/\sqrt{\Phi}.
\end{equation}
Since $\Phi$ is larger when the particles are trapped than when they are untrapped, we conclude that, unless $T_w$ drastically changes during trapping, Eq.~(\ref{A2}) is the only necessary condition for adiabaticity. 

At this stage we must note that, for our derivation to be complete, we need to  account for the divergence of the frozen period close to the separatrix. To do so, we choose for the normalizing velocity scale, $v_{th}=p_0/M-V_\phi$, where $V_\Phi$ is the value assumed by the wave phase velocity when the particles orbits are close to the separatrix. Then, close to the separatrix, $v_0-v_\phi =1$ and $\varepsilon=1/T_w$. From neo-adiabatic theory, we know that only the values of the period $T$ corresponding to $h\equiv H-\Phi \sim \varepsilon$ really matter. When $\varepsilon \rightarrow 0$, this period diverges as $\ln(\varepsilon)$, so that $T/T_w \sim \varepsilon \ln(\varepsilon)$. Therefore, in spite of the divergence of $T$, $T/T_w$ remains  small when $\varepsilon \ll 1$, and neo-adiabatic results do apply. \\

Let us now address the situation when the initial velocity distribution function, $f_0$, is smooth. In this situation,  Eq.~(\ref{A2}) cannot be satisfied for all values of $v_0$ since, usually, for any $v_\phi$ there exists particles such that $v_0 = v_\phi$. Now, let $v_T$ be the typical velocity range of variation of $f_0$ about $v_\phi$. If 
\begin{equation}
\label{A4}
\varepsilon_S \equiv 1/v_TT_w \ll1,
\end{equation}
the weight of non-adiabatic particles, which do not fulfill Eq.~(\ref{A2}), is small and, in many instances, negligible (see the discussion below regarding the need to introduce an intermediate time scale related to $1/\sqrt{\Phi}$). Consequently, for a smooth initial distribution function, Eq.~(\ref{A4}) may be viewed as the condition for adiabaticity. \\

For the Hamiltonian $H_2$, defined by Eq.~(\ref{H2}), the same conclusions hold, except that $T_w$ which is still associated to the time variations of the wave properties, now corresponds the typical time of variation of $\Phi$ of of $\dot{v}_\phi$ (and not of $v_\phi$). \\

To be complete, we now need to mention an intermediate time scale related to $1/\sqrt{\Phi}$. Discussing this issue in detail would actually be quite complicated, and outside the scope of the paper. The importance of this time scale depends on the particular physical problem that one wants to address. In a sense, the need to introduce this time scale amounts to discussing when Eq.~(\ref{A4}) indeed guarantees that one may use the adiabatic approximation for a smooth initial distribution function.  Here, we will only quickly mention two examples.

In the companion paper~\cite{companion}, we show that the adiabatic distribution function provides an accurate estimate of the frequency of an electron plasma wave (EPW), regardless of how small $\Phi$ is. Hence, as regards the derivation of the linear, or nonlinear, EPW dispersion relation, there is no relevant time scale related to $1/\sqrt{\Phi}$. Eq.~(\ref{A4}) is enough. 

By contrast, one cannot address the collisionless dissipation of a small amplitude EPW by making use   of the adiabatic approximation, because one would miss Landau damping. As discussed in several papers~\cite{benisti07,dissipation,benisti16}, provided that Eq.~(\ref{A4}) is fulfilled, the adiabatic distribution function allows to accurately derive the EPW collisionless dissipation whenever $\int \sqrt{\Phi} d\tau >2\pi$, hence after a time of the order of $1/\sqrt{\Phi}$. 

\section{Relation between the condition for adiabaticity and the ratio $\dot{v}_\phi/\Phi$}
\label{A}
\setcounter{equation}{0}
\setcounter{app}{2}
For the sake of simplicity, in order to discuss the condition on $\dot{v}_\phi/\Phi$ for the dynamics to be indeed adiabatic, we choose a nearly constant rate of variation for the wave amplitude and phase velocity. Namely we choose,
\begin{eqnarray}
\label{73}
\Phi& =&\Phi_0 \left[e^{\gamma_1 t}-1\right], \\
\label{74}
v_\phi &=&v_{\phi_0}  \left[e^{\gamma_2 t}-1\right].
\end{eqnarray}
Moreover, we consider the situation when all the particles have the same initial velocity, $v_0>0$, and we also impose, 
\begin{eqnarray}
\label{80}
\gamma_1&=&\frac{\sqrt{2}\gamma_2}{\varepsilon_2} \\
 \label{81}
 v_0 &=& \frac{16 \gamma_2}{(\sqrt{8}-2) \varepsilon \pi^2},\\
 \label{82}
v_{\phi_0} &=& \frac{8 \gamma_2}{  \varepsilon \pi^2} \left(\frac{\varepsilon^2 \pi^2 \Phi_0}{8 \gamma_2^2} \right)^{\varepsilon_2/\sqrt{2}}.
\end{eqnarray}
Then, when separatrix crossing occurs i.e., when $v_0-v_\phi \approx v_{tr}$, $v_0/\gamma_2$ is maximum and, 
\begin{eqnarray}
\label{78}
\dot{v}_\phi &=& \varepsilon \Phi, \\
\label{79}
\dot{v}_\phi &=& \varepsilon_2 \dot{v}_{tr}.
\end{eqnarray}
From Eqs.~(\ref{79})~and~(\ref{68}), one changes the trapping probability by varying $\varepsilon_2$. As for Eq.~(\ref{78}), it directly relates $\dot{v}_\phi/\Phi$ to $v_0/\gamma_2$. Indeed, from Eqs.~(\ref{81}) and~(\ref{78}), 
\begin{equation}
\label{n10}
\frac{\dot{v}_\phi}{\Phi}=\frac{16 \gamma_2}{(\sqrt{8}-2)} \frac{\gamma_2}{v_0}.
\end{equation}

 Now, using the definition of Appendix A, $\gamma_2 \sim 1/T_w$. Moreover, since $v_\phi(0)=0$, $v_0 \sim v_0-v_\phi$. Actually, for the example chosen in this Appendix, $\vert v_0-v_\phi \vert < v_0$. Hence, we conclude from Eq.~(\ref{n10}) that, when $\dot{v}_\phi/\Phi$ is not small, $\varepsilon \equiv 1/\vert v_0-v_\phi \vert T_w$ is not small either, so that the condition for adiabaticity, Eq.~(\ref{A2}), is not fulfilled. 

\section{Condition for no trapping}
\label{B}
\begin{figure}[!h]
\centerline{\includegraphics[width=12cm]{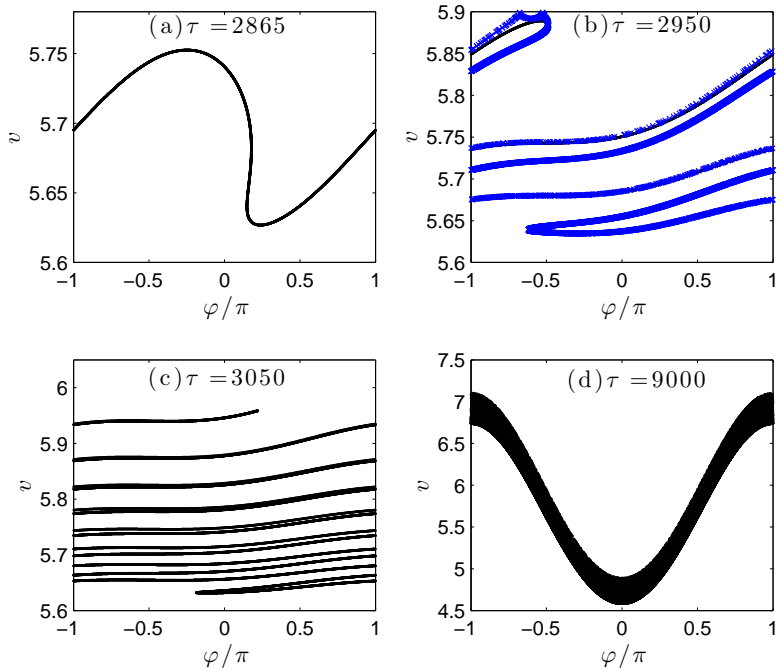}}
\caption{\label{f0} (Color online) The black curves are obtained by numerically solving the equations of motion for 65536 particles having all the same initial velocity, $v_0=5.7$, and whose initial positions are uniformly distributed between $-\pi$ and $+\pi$. Moreover, $\Phi=\Phi_0 \exp[\gamma_0\sin(\gamma_1 \tau)]$ and $v_\phi=\gamma_2 \tau$, with $\phi_0=10^{-5}$, $\gamma_0=18$, $\gamma_1=10^{-4}$ and $\gamma_2=2\times10^{-3}$. The blue crosses in Fig.~\ref{f0}(b) are obtained by solving $H_2=H_2(\tau=0)$. } 
\end{figure}

\setcounter{app}{3}
When $\dot{v}_\phi > \Phi$, there is no separatrix for the Hamiltonian $H_2$ defined by Eq.~(\ref{H2}) and, therefore, using $H_2$ one would conclude that trapping cannot occur.  By contrast, a separatrix always exists for $H$ defined by Eq.~(\ref{H}), so that trapping is expected whenever $H<\Phi$. Now, the notion of trapping is actually not clear for a non-constant Hamiltonian. Indeed, usually a particle is said to be trapped when its normalized position, $\varphi$, can \textit{never}~change by more than $2\pi$. When the Hamiltonian is time-dependent, such a condition cannot be guaranteed for all times, and the definition of trapping should, therefore, be slightly modified. Henceforth, a particle will be considered as trapped provided that $\varphi$ has experienced, at least, one whole oscillation within an interval whose extent is less than $2\pi$. The period of such an oscillation is of the order of $T_B \equiv 2\pi/\sqrt{\Phi}$, which is large when $\dot{v}_\phi>\Phi$ (since, by hypothesis, $v_\phi$ varies slowly in time). Then, when $\dot{v}_\phi$ remains nearly constant, $H_2$ is better conserved than $H$. In this situation, the conclusions drawn from $H_2$ should be more accurate than those obtained with $H$. In particular, the no-trapping prediction when $\dot{v}_\phi >\Phi$ should be correct. We tested this numerically, and we always found that no trapping occurred if $\dot{v}_\phi/\Phi$ was larger than a value close to unity when Eq.~(\ref{49})~or~(\ref{50}) were fulfilled, as may be seen in Fig.~\ref{f0}.  

This figure also shows that, when $\dot{v}_\phi/\Phi>1$, the orbits found by numerically solving the equations of motion are not $2\pi$-periodic in $\varphi$. Actually, if one plots $v$ as a function of $\varphi [2\pi]$, so  that $-\pi \leq \varphi \leq \pi$, the orbits exhibit an increasing number of foldings, which are illustrated in Fig.~\ref{f0}. This is because these orbits are not solutions of $H \approx Const.$ As is clear from  Fig.~\ref{f0}(b), they are solutions of $H_2 \approx Const.$, and $H_2$ is not periodic in $\varphi$. If the wave keeps growing and $\dot{v}_\phi/\Phi \ll 1$, the orbits look again $2\pi$-periodic, as may be seen in Fig.~\ref{f0}(d), and the frozen orbits  of $H$ and $H_2$ are actually close to each other. However, the curve $v(\varphi)$ in Fig.~\ref{f0}(d) is significantly thicker than in Fig.~\ref{f0}(a). This is due to the multiple foldings which occurred earlier and, clearly, the thickness of the curve in Fid.~\ref{f0}(d) is close the interval spanned by $v$ in Figs.~\ref{f0}(b) and \ref{f0}(c). 

Note also that, for the parameters chosen for Fig.~\ref{f0} when the initial particles velocity is $v_0=5.7$, no trapping is observed numerically while, when $I-v_\phi=v_{tr}$, $\dot{v}_\phi/\Phi \approx 1.3$. If  we choose $v_0=5.8$ instead of $v_0=5.7$, we numerically find that some particles (about 2.5\% of them) get trapped while, when $I-v_\phi=v_{tr}$, $\dot{v}_\phi/\Phi \approx 1.2$. Hence, the no-trapping condition,  $\dot{v}_\phi/\Phi >1$ is, indeed, accurate. Note also that, the trapping probability derived using $H$, and given by Eq.~(\ref{68}) of Paragraph~\ref{III.1}, is $p^{(\alpha \rightarrow \gamma)} \approx 0.13\%$. Hence, it is very small anyway but, using 65536 initial conditions numerically, we would have found trapped particles if Eq.~(\ref{68}) was accurate. 

%\section{Distribution in $h$ for an initial Dirac distribution in action}
%\label{D}
%\setcounter{app}{4}


\begin{thebibliography}{10}
\bibitem{note}
Within the classical adiabatic theory, one may define better conserved invariants than the action, that would only vary by a term proportional to $\varepsilon^n$ with $n>1$~\cite{lenard,nei}. Indeed, adiabatic theory is nothing but a perturbative expansion in the slowness of the dynamics, that may be led to any desired order. However, it is well-known that such an expansion would break down near the separatrix~\cite{cary}. Consequently, introducing higher order invariants would not be useful in this paper, which is devoted to the change in the adiabatic invariant due to separatrix crossing. Indeed, from neo-adiabatic theory we know that separatrix crossing entails a change in the adiabatic invariant by a term proportional to $\varepsilon \ln(\varepsilon)$, whether the invariant is defined from the action, as done in this paper, or from a higher order term in the classical adiabatic expansion.
\end{thebibliography}
\end{document}